\documentstyle[psfig]{mn}

\newcommand{\vi}{\vec{v}}
\newcommand{\vxi}{\vec{\xi}}
\newcommand{\lp}{ \left(}
\newcommand{\rp}{ \right)}
\newcommand{\na}{ \vec{\nabla} }
\newcommand{\disp}[1]{\displaystyle #1}
\newcommand{\er}{\vec{e}_r}
\newcommand{\noi}{ \noindent }
\newcommand{\dnr}[1]{\frac{d  #1}{dr}}
\newcommand{\ddnr}[1]{\frac{d^2  #1}{dr^2}}
\def\Ln{\mathop{\hbox{ln}}\nolimits}
\newcommand{\beq}{\begin{equation}}
\newcommand{\eeq}{\end{equation}}
\newcommand{\dr}[1]{\frac{\partial  #1}{\partial r}}
\newcommand{\ltex}[1]{\quad \hbox{#1} \quad}
\newcommand{\llp}{ \ell(\ell+1)}
\newcommand{\lc}{ \left[}
\newcommand{\rc}{ \right]}
\newcommand{\greq}{\begin{equation}\left\{ \begin{array}{l}}
\newcommand{\egreq}{\end{array}\right. \end{equation}}
\newcommand{\uu}{ u^\ell_m }
\newcommand{\vv}{ v^\ell_m }
\newcommand{\YL}{ Y^m_\ell }
\newcommand{\dnx}[1]{\frac{d  #1}{dx}}
\newcommand{\eeqn}[1]{\label{#1}\end{equation}}
\newcommand{\od}[1]{\mbox{${\cal O}(#1)$}}
\newcommand{\eq}[1]{(\ref{#1})}
\renewcommand{\theenumi}{(\arabic{enumi})}

\title[The anelastic and subseismic approximations for low-frequency
g-modes]{A comparison of the anelastic and subseismic approximations
for low-frequency gravity modes in stars}

\author[B. Dintrans and M. Rieutord]
{B. Dintrans$^{1,2}$ and M. Rieutord$^{2,3}$\\
$^1$Nordic Institute for Theoretical Physics, Blegdamsvej 17, DK-2100
Copenhagen, Denmark\\
$^2$Laboratoire d'Astrophysique de Toulouse, Observatoire
Midi-Pyr\'en\'ees, 14 avenue E. Belin, 31400 Toulouse, France\\
$^3$Institut Universitaire de France}

\pagerange{\pageref{firstpage}--\pageref{lastpage}}
\pubyear{2000}

\begin{document}

\maketitle

\label{firstpage}

\begin{abstract}
A comparative study of the properties of the anelastic and subseismic
approximations is presented. The anelastic approximation is commonly
used in astrophysics in compressible convection studies whereas the
subseismic approximation comes from geophysics where it is used to
describe long-period seismic oscillations propagating in the Earth's
outer fluid core. Both approximations aim at filtering out the acoustic
waves while retaining the density variations of the equilibrium
configuration. However, they disagree on the form of the equation of
mass conservation. We show here that the anelastic approximation is in
fact the only consistent approximation as far as stellar low-frequency
oscillations are concerned.  We also show that this approximation
implies Cowling's approximation which neglects perturbations of the
gravity field. Examples are given to assert the efficiency of the
anelastic approximation.  \end{abstract}


\section{Introduction}

Initiated with the work on atmospheric convection by Ogura \& Phillips
(1962) and Gough (1969), the anelastic approximation has been widely
used in astrophysics to describe compressible convection.  For example,
it has been first applied by Latour et al. (1976) and Toomre et al.
(1976) to investigate the non-rotating and non-magnetized convective
envelopes. Then, Gilman \& Glatzmaier (1981) and Glatzmaier \& Gilman
(1981a,b,c) extended these studies to the rotating case whereas the MHD
case has been considered by Glatzmaier (1984,1985a,b), Lantz \& Sudan
(1995) and Lantz \& Fan (1999).

In all these studies, this approximation has been preferred to an other
famous one in fluid dynamics, the so-called Boussinesq approximation
where the velocity is assumed to be solenoidal, i.e. $\na\cdot \vi = 0$
(Spiegel \& Veronis 1960). As the density varies on several orders of
magnitude in a
convection zone, using the anelastic approximation is indeed more
relevant and leads to $\na\cdot (\rho \vi) = 0$. However, both
approximations filter out short-period acoustic oscillations by
assuming that the Mach number of the convection is small so that `the
task of numerical solution will not be complicated by the need to
resolve very rapid time variations' (Latour et al. 1976).

More recently, the anelastic approximation has been used in the field
of stellar oscillations. Waves can propagate over long distances in
star interiors and large variations of density need also to be taken
into account. Thus, as in stellar convection studies, the Boussinesq
approximation is too restrictive and not adapted for this problem.
However, since acoustic waves are filtered out, only the low-frequency
part of the oscillations spectrum is captured. As this part of the
spectrum also corresponds to the most perturbed one when rotation acts,
this approximation is very attractive when low-frequency modes of a
rotating star are considered.

Understanding the spectrum of rotating stars is indeed a difficult problem
when the rotation period is of the same order than the oscillation
one. In this case, the usual perturbative theory of Ledoux (1951)
reaches its limits and the eigenvalue problem needs to be solved
non-perturbatively (Dintrans \& Rieutord 2000, hereafter referred to as
Paper I). But rotation introduces new computational challenges: ({\em
i}) the strong Coriolis coupling between normal modes of different
degrees $(\ell,\ell\pm 1)$ leads to a large system of coupled
differential equations; ($ii$) some singularities due to the presence
of wave attractors emerge in the region $\sigma \simeq 2 \Omega$,
$\sigma$ being the wave frequency and $\Omega$ the uniform rotation
rate of the star. Applying the anelastic approximation does not remove
these singularities but decreases the size of the numerical problem
since acoustic quantities disappear (see Paper I). We note that
Berthomieu et al. (1978) used the same arguments when considering a
hierarchy between the displacement and pressure components, this
hierarchy leading to the same result as the direct use of the anelastic
approximation.

Contrary to stars, the mass of the Earth is not strongly concentrated
near its centre. Only 30\% of the whole mass is indeed inside the half
radius whereas this ratio is around 90\% for the Sun (see e.g. Stix
1989). It means that the variations in density in the Earth's outer
fluid core generated by the propagation of seismic P-waves may produce
non negligible variations in the gravitational potential $\phi$ and
explains why geophysicists do not wish to use the Cowling approximation
in their description of P-waves. To take into account this effect but
to simplify the equations for low frequency modes Smylie \& Rochester
(1981), Smylie, Szeto \& Rochester (1984) introduced the `subseismic'
approximation where self-gravity is included.

Recently, De Boek, Van Hoolst \& Smeyers (1992) applied this
approximation to describe the low-frequency g-modes of non-rotating
stars. Starting from the same equations as Smylie \& Rochester (1981),
they found an analytic expression for the {\em asymptotic} low
frequencies which slightly differs from the standard one deduced by
Tassoul (1980) in the Cowling approximation. By applying this
expression to the low-frequency oscillations of both homogeneous and
polytropic ($n=3$) star models, De Boek et al. (1992) concluded that
the subseismic approximation is more accurate than the standard one in
the polytropic case but less accurate in the homogeneous case.

However, as will be shown below, the subseismic approximation is not a
consistent approximation of the equations of motion; we will show indeed
that the proper way to simplify the equations in order to deal with the
low frequency modes is to use the anelastic approximation and that this
approximation also implies Cowling's approximation.

Hence, after establishing the complete set of equations describing linear
oscillations of a stratified compressible fluid and the appropriate
boundary conditions (Section 2), we focus on the similarities and
differences of the anelastic and subseismic approximations (Section 3);
in particular, we show that both imply Cowling's approximation and
demonstrate the inconsistency of the subseismic approximation. Then
we compare the two approximations on the test cases used by De Boek et
al. (1992), namely, the homogeneous and polytropic star models (Section
4). Comparison of the resulting approximate eigenfrequencies with the
exact ones (computed numerically for the polytrope but analytically
for the homogeneous model) shows clearly the better efficiency of the
anelastic approximation. Finally, we conclude in Section 5 with some
outlooks of our results.

\section{The basic equations}

Assuming a time-dependence of the form $\exp (i \sigma t)$, the
governing equations that describe the adiabatic oscillations of a
spherically non-rotating star are given by

\begin{eqnarray}
& & \rho' + \na\cdot ( \rho \vxi ) = 0, \label{eq1} \\ \nonumber \\
& & \disp \sigma^2 \vxi = \na \lp \frac{P'}{\rho} + \phi' \rp - 
\frac{N^2}{\rho g} \delta P \er, \label{eq2} \\ \nonumber \\
& & \disp \delta P = c^2 \delta \rho, \label{eq3} \\ \nonumber \\
& & \nabla^2 \phi' = 4 \pi G \rho', \label{eq4}
\end{eqnarray}

\noi where $P',\rho'$ and $\phi'$ respectively denote the Eulerian
fluctuations of pressure, density and gravitational potential whereas
$\delta P,\delta \rho$ and $\vxi$ are the Lagrangian fluctuations
of pressure, density and displacement such as

\[
\delta P = P' + \dnr{P} \xi_r, \qquad \delta \rho = \rho' + \dnr{\rho} 
\xi_r,
\]

\noi with a pressure gradient satisfying the hydrostatic equilibrium
$dP/dr=-\rho g$. Also, $\rho$ is the equilibrium density, $\vec{g}$ the
gravity and $\gamma=(\partial \Ln P / \partial \Ln \rho)_S$ the first
adiabatic exponent. Finally, $c^2$ and $N^2$ respectively denote the
squares of sound speed velocity and Brunt-V\"ais\"al\"a frequency such
as

\beq
c^2 = \gamma \frac{P}{\rho}, \qquad N^2=g \lp \frac{1}{\gamma} 
\dnr{\Ln P} - \dnr{\Ln \rho} \rp.
\eeqn{def_bv}

\noi In order to obtain a well-posed eigenvalue problem with an eigenvalue
$\sigma^2$, boundary conditions should be stipulated at the star centre
and surface. At the centre, we impose the regularity of all pulsation
quantities such as (see e.g. Unno et al. 1989)

\beq
\xi_r \propto r^{\ell-1}, \qquad P' \propto r^\ell, \qquad \phi' \propto
r^\ell,
\eeq

\noi where $\ell$ denotes the spherical harmonics degree of the
eigenmode. At the surface, which is assumed to be free, we impose the
following condition coupling $\xi_r$ and $\phi'$ (Ledoux and Walraven
1958)

\beq
\lp \dnr{\phi'} \rp_R + \frac{\ell+1}{R} \phi' (R) = - 4 \pi G (\rho
\xi_r)_R.
\eeq

\noi Finally, we adopt the classical mechanical outer condition for the
pressure which reads

\beq
\delta P = 0 \ltex{at} r=R.
\eeq

\section{The subseismic and anelastic approximations applied to stellar
oscillations}

\subsection{The common properties}

Since the work of Cowling (1941), it is well known that g-modes are
mainly transverse whereas p-modes are predominantly radial. The
Eulerian pressure and density perturbations are thus small for a g-mode
and both subseismic and anelastic approximations take advantage of that
by assuming that the Lagrangian pressure fluctuation $\delta P$ is only
due to the radial displacement one; i.e. the Eulerian fluctuation $P'$
does not contribute and we have

\beq
\delta P = P' + \dnr{P} \xi_r \simeq \dnr{P} \xi_r = - \rho g \xi_r.
\eeqn{eq0}

\noi In this case, Eqs.~\eq{eq2}, \eq{eq3} and \eq{eq4} read now

\begin{eqnarray}
& & \disp \sigma^2 \vxi = \na \lp \frac{P'}{\rho} + \phi' \rp + 
N^2 \xi_r \er, \label{eq5} \\ \nonumber \\
& & \rho' = \frac{N^2}{g} \rho \xi_r, \label{eq6} \\ \nonumber \\
& & \nabla^2 \phi' = 4 \pi G \frac{N^2}{g} \xi_r.
\end{eqnarray}

\noi At this stage, two important consequences appear:

\begin{enumerate}

\item As observed by De Boek et al. (1992), the radial Lagrangian
displacement necessarily vanishes at the surface. To show this, we take
the radial component of Eq.~\eq{eq5}

\[
\sigma^2 \xi_r = \dr{} \lp \frac{P'}{\rho} + \phi' \rp + N^2 \xi_r.
\]

\noi The problem arises from the fact that the Brunt-V\"ais\"al\"a
frequency diverges as $1/c^2$ at the star surface (see
Eq.~\ref{def_bv}). Thus, as $N^2$ diverges to infinity at $r=R$,
$\xi_r$ must vanish at this point to avoid a singular radial
displacement. In the complete case (i.e. when acoustic waves are
included), we note that it is the mechanical condition $\delta P = 0$
which permits a finite radial displacement at the surface, as shown by
the radial component of Eq.~\eq{eq2}

\[
\sigma^2 \xi_r = \dr{} \lp \frac{P'}{\rho} + \phi' \rp - \frac{N^2}{\rho
g} \delta P,
\]

\noi where the term $N^2/ (\rho g)$ diverges as $1/ (\rho c^2)$. Thus, the
subseismic and anelastic $\xi_r$-eigenfunctions always have one node less
than their corresponding unapproximated eigenfunctions for which
$\xi_r(R)$ is finite.

\item The second consequence, which is not straightforward, is that the
use of the subseismic or anelastic approximation necessarily implies
Cowling's approximation where the perturbations $\phi'$ are neglected
(Cowling 1941). To show this, we first take the curl of Eq.~\eq{eq5}
and obtain

\[
\sigma^2 \na\times \vxi = \na\times (N^2 \xi_r \er).
\]

\noi Furthermore, combining Eqs.~\eq{eq1} and \eq{eq6} allows us to
find the following subseismic form for the equation of mass conservation

\beq
\frac{N^2}{g} \rho \xi_r + \na\cdot (\rho \vxi) = 0 \quad\Rightarrow
\quad\na\cdot \vxi = \frac{g}{c^2} \xi_r,
\eeqn{eq_sub}

\noi whereas, as shown in the introduction, the anelastic counterpart
of this equation reads

\beq
\na\cdot (\rho \vxi) = 0 \quad\Rightarrow\quad \na\cdot \vxi = - \dnr{\Ln \rho} 
\xi_r.
\eeqn{eq_anel}

\noi The two different forms \eq{eq_sub} and \eq{eq_anel} may be
formally written ${\cal L} (\vxi) = 0$ where $\cal L$ denotes a
differential operator depending on the approximation we use. We have
thus to deal with the new following system

\begin{eqnarray}
& & \sigma^2 \na\times \vxi = \na\times (N^2 \xi_r \er), \label{eq7} 
\\ \nonumber \\
& & {\cal L} (\vxi) = 0, \label{eq8} \\ \nonumber \\
& & \nabla^2 \phi' = 4 \pi G \frac{N^2}{g} \xi_r, \label{eq9}
\end{eqnarray}

\noi whereas the surface boundary conditions now read 

\[
\xi_r (R) = 0 \ltex{and} \lp \dnr{\phi'} \rp_R + \frac{\ell+1}{R} \phi'
(R) = 0.
\]

It is thus clear that the Poisson equation \eq{eq9} decouples from
Eqs.~\eq{eq7} and \eq{eq8}. In other words, the knowledge of 
Eqs.~\eq{eq7} and \eq{eq8} with the boundary condition $\xi_r(R)=0$ is
sufficient to find the set of eigenfrequencies $\sigma^2$.  The
gravitational potential perturbation $\phi'$ does not act on the
determination of $\sigma^2$ and can be seen as `slaved' to the $\xi_r$
via Eq.~\eq{eq9}\footnote{It explains why Smeyers et al.  (1995),
starting from the complete case, found the same asymptotic development
of $\sigma^2$ as those derived by Tassoul (1980) under the Cowling
approximation.}. We note that Smylie \& Rochester (1981) do not clearly
emphasize this point and just noticed that the $\phi'$-perturbations
can be determined {\em after} solving their subseismic equations
for velocity and reduced pressure.

\end{enumerate}

\subsection{The main difference}

As shown above, the subseismic and anelastic approximations do not lead
to the same equation of mass conservation since the term stemming from
the density perturbations is kept in the subseismic case (see
Eqs.~\ref{eq_sub} and \ref{eq_anel}). We will now show that the
subseismic form of this equation is in fact not correct because it is
not consistent with the basic assumption from Eq.~\eq{eq0}.

First of all, we recall that both approximations assume that the Eulerian
pressure perturbation does not contribute to the Lagrangian one, that is

\beq
\frac{P'}{P} \ll - \xi_r \dnr{\Ln P}\quad \Leftrightarrow
\quad\frac{P'}{P} \ll \frac{\xi_r}{H_p},
\eeqn{ineq_P}

\noi where $H_p = (-d\Ln P /dr)^{-1}$ denotes the pressure scale
height. Moreover, we can relate the Eulerian pressure and density
perturbations using Eq.~\eq{eq3} as

\[
\frac{P'}{P} = \gamma \lp \frac{\rho'}{\rho} - \frac{N^2}{g} \xi_r \rp,
\qquad \frac{N^2}{g} = \frac{1}{H_\rho} - \frac{1}{\gamma H_p},
\]

\noi where $H_\rho = (-d\Ln \rho /dr)^{-1}$ denotes the density scale
height and where Eq.~\eq{def_bv} has also been used. Therefore the basic
assumption from Eq.~\eq{ineq_P} leads to

\beq
\frac{\rho'}{\rho} - \lp \frac{1}{H_\rho} - \frac{1}{\gamma H_p} \rp \xi_r
\ll \frac{\xi_r}{\gamma H_p}\quad \Rightarrow\quad \frac{\rho'}{\rho} \ll
\frac{\xi_r}{H_\rho}.
\eeqn{ineq_rho}

\noi This last equation is important since it shows that we can also
neglect the Eulerian contribution $\rho'$ in the Lagrangian
perturbation $\delta \rho$ when Eq.~\eq{eq0} is assumed, i.e. we have

\[
\delta \rho = \rho' + \xi_r \dnr{\rho} \simeq \xi_r \dnr{\rho}.
\]

\noi In other words, if we just take into account the contribution
stemming from the equilibrium pressure gradient in $\delta P$, we
should also do the same with the contribution coming from the
equilibrium density gradient in $\delta \rho$. Hence

\beq
\delta P \simeq \xi_r \dnr{P} \ltex{is equivalent to} \delta \rho
\simeq \xi_r \dnr{\rho}.
\eeqn{full.approx}

We can now easily understand why the subseismic equation of mass
conservation is not correct. Eq.~\eq{eq1} indeed gives

\[
\rho' + \na\cdot (\rho \vxi) = 0 \quad\Rightarrow\quad \delta \rho + 
\rho \na\cdot \vxi = 0.
\]

\noi In order to be consistent with the basic assumption from 
Eq.~\eq{full.approx}, the equation of mass conservation now reads

\[
\xi_r \dnr{\rho} + \rho \na\cdot \vxi = 0 \quad\Rightarrow\quad 
\na\cdot \vxi = - \dnr{\Ln \rho} \xi_r,
\]

\noi and we recover the good anelastic form \eq{eq_anel}.

The subseismic form of mass conservation \eq{eq_sub} of Smylie \&
Rochester (1981) is therefore clearly inconsistent because of its
incompatibility with the hypothesis \eq{eq0}. As a consequence, the
subseismic approximation is necessarily less accurate than the
anelastic one as we will now show.

\section{Results}

In order to test the efficiency of both approximations, we applied the
following procedure to the oscillations of the homogeneous compressible
star (i.e. a polytrope of order 0) and the polytrope $n=3$:

\begin{enumerate}

\item we first calculated the exact eigenfrequencies of the full case,
that is we solved the set of equations (1-4) using boundary conditions
(6-8). This has been achieved analytically for the polytrope $n=0$
(Pekeris 1938) and numerically for the polytrope $n=3$.

\item we next calculated the subseismic and anelastic eigenfrequencies
by solving the system consisting in Eq. \eq{eq7}, the equation of
mass conservation \eq{eq_sub} or \eq{eq_anel} and finally the boundary
condition $\xi_r(R)=0$ discussed in Section 3.1. Once again, analytical
expressions have been used in the homogeneous case whereas the polytrope
$n=3$ has been computed numerically. The efficiency of both
approximations has been deduced by a direct comparison between
these approximated eigenfrequencies and the previous exact ones.

\end{enumerate}

\subsection{The homogeneous polytrope}

As a first example, we concentrate on the unstable second-degree
g$^-$-modes of the polytrope $n=0$. Pekeris (1938) first studied the
nonradial oscillations of this model and found the exact
g$^-$-eigenfrequencies as

\[
\omega^2 = \Delta_{\ell k} - \sqrt{\Delta^2_{\ell k} + \Lambda},
\qquad (k=0,1,2,\dots),
\]

\noi where $\Lambda = \llp$ and

\[
\Delta_{\ell k} = \gamma \lc k \lp \ell+k+\frac{5}{2} \rp + \ell +
\frac{3}{2} \rc - 2.
\]

\noi Here $\omega^2$ denote the dimensionless eigenfrequencies, which
for a star of mass $M$ and radius $R$, are related to the
$\sigma^2$-ones by

\[
\sigma^2 = \frac{GM}{R^3} \omega^2.
\]

Using an equivalent method as the one used by Pekeris (i.e. the
search of a power-serie solution), we calculated the analytic
expressions of the subseismic and anelastic eigenfrequencies of the
homogeneous polytrope and found (see appendix A)

\begin{eqnarray}
& & \disp \omega^2_{\hbox{\scriptsize sub}} = \frac{2}{\gamma} 
\frac{\Lambda}{\Lambda - (\ell+2k')\lp \ell + 2k' +1 + \frac{2}{\gamma} 
\rp}, \label{w_sub} \\ \nonumber \\
& & \disp \omega^2_{\hbox{\scriptsize anel}} = \frac{2}{\gamma} 
\frac{\Lambda}{\Lambda - (\ell+2k')(\ell+2k'+1)}, \label{w_anel}
\end{eqnarray}

\noi where $(k'=1,2,\dots)$. 

\begin{table}
\caption{Squares of exact $\omega^2$, subseismic
$\omega^2_{\hbox{\scriptsize sub}}$ and anelastic $\omega^2_{\hbox{
\scriptsize anel}}$ eigenfrequencies of the homogeneous polytrope.}
\begin{tabular}{lccc}
\hline
& $\omega^2 \times 10^2$ & $\omega^2_{\hbox{\scriptsize sub}} \times 10^2$ & 
$\omega^2_{\hbox{\scriptsize anel}} \times 10^2$ \\ \hline
g$^-_3$    & -7.251703 & -6.206896 & -6.923077 \\
g$^-_5$    & -3.613671 & -3.260869 & -3.529411 \\
g$^-_{10}$ & -1.221965 & -1.156069 & -1.212121 \\
g$^-_{15}$ & -0.610579 & -0.587851 & -0.608108 \\
g$^-_{20}$ & -0.365629 & -0.355239 & -0.364741 \\
g$^-_{25}$ & -0.243308 & -0.237718 & -0.242914 \\
g$^-_{30}$ & -0.173527 & -0.170180 & -0.173327 \\ \hline
\end{tabular}
\end{table}

\begin{figure}
\psfig{file=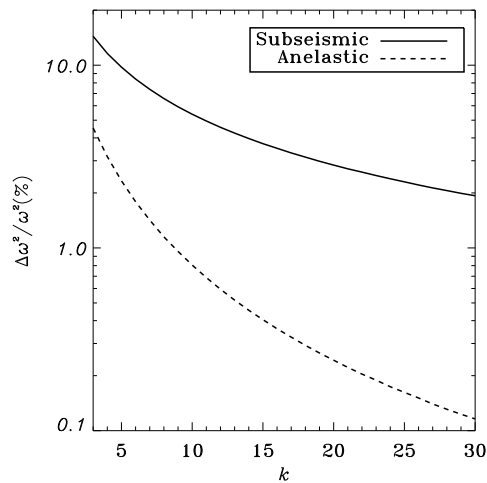,height=7cm}
\caption{Rates of relative errors of the subseismic (Eq.~\ref{w_sub}) 
and anelastic (Eq.~\ref{w_anel}) eigenfrequencies as the
order $k$ increases for the homogeneous star model.}
\label{fig1}
\end{figure}

We summarized in Table 1 the exact eigenfrequencies calculated for the
homogeneous star with $\gamma = 5/3$, $\ell=2$ and various orders $k$,
whereas Fig.~\ref{fig1} gives the relative errors (in percents) made
with both approximations for the first thirty g$^-$-modes.  The
anelastic approximation clearly appears to be more precise than the
subseismic one since, for example, the anelastic eigenfrequency at
$k=30$ is about twenty times more precise than the subseismic one.

\subsection{The polytrope $n=3$}

\subsubsection{Numerics}

We next calculated the second-degree g$^+$-mode of the polytrope $n=3$.
As already mentioned, analytic expressions of eigenfrequencies are not
known for this model then we computed numerically the eigenfrequencies
of g$^+$-modes with the complete set of equations and their
approximated subseismic and anelastic counterparts. As in our preceding
work (Dintrans, Rieutord \& Valdettaro 1999; paper I), we used an
iterative eigenvalue solver based on the incomplete Arnoldi-Chebyshev
algorithm.  This numerical method differs from the classical ones where
eigenfrequencies are determined by a direct integration of the
governing equations using either relaxation methods (Osaki 1975) or
shooting methods (Hansen \& Kawaler 1994). Our eigenvalue formulation
is presented in appendix B, with a discussion on the tricky problem of
the degeneracy of the eigenvalue equations at the star surface when the
subseismic or anelastic approximation are used.

Eigenfrequencies $\lambda^2$ are solutions of a generalized eigenvalue
problem

\[
{\cal M}_A \vec{\Xi} = \lambda^2 {\cal M}_B \vec{\Xi},
\]

\noi where ${\cal M}_A,{\cal M}_B$ denote two differential operators
discretized on a Gauss-Lobatto grid associated with Chebyshev polynomials
and $\vec{\Xi}$ is the eigenvector associated with $\lambda^2$. We also
note that the regular singularities appearing in the subseismic or
anelastic equations at the surface are replaced by the boundary
conditions (see appendix B.2).

\subsubsection{Results}

\begin{table}
\caption{Squares of the true $\lambda^2$, subseismic
$\lambda^2_{\hbox{\scriptsize sub}}$ and anelastic
$\lambda^2_{\hbox{\scriptsize anel}}$ eigenfrequencies of the polytrope
$n=3$.}
\begin{tabular}{lccc}
\hline
& $\lambda^2 \times 10^2 $ & $\lambda^2_{\hbox{\scriptsize sub}} \times 10^2$ &
$\lambda^2_{\hbox{\scriptsize anel}} \times 10^2 $ \\ \hline
g$_3$    & 1.121064 & 1.269076 & 1.210566 \\
g$_5$    & 0.576059 & 0.627794 & 0.603954 \\
g$_{10}$ & 0.198396 & 0.208345 & 0.202646 \\
g$_{15}$ & 0.099755 & 0.103233 & 0.100973 \\
g$_{20}$ & 0.059984 & 0.061587 & 0.060453 \\
g$_{25}$ & 0.040039 & 0.040911 & 0.040259 \\
g$_{30}$ & 0.028631 & 0.029154 & 0.028744 \\ \hline
\end{tabular}
\end{table}

\begin{figure}
\psfig{file=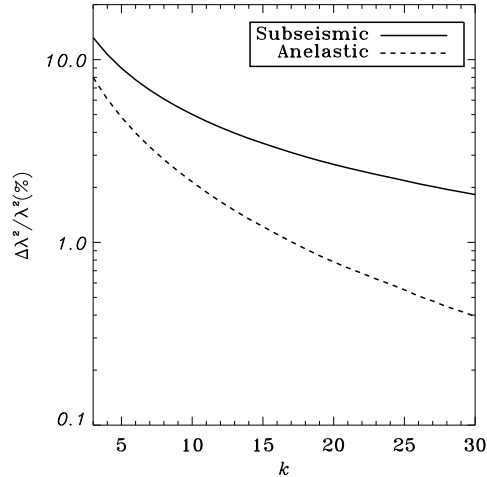,height=7cm}
\caption{Rates of relative errors of computed eigenfrequencies for the 
polytrope $n=3$.}
\label{fig2}
\end{figure}

We summarized in Table 2 the computed eigenfrequencies (in units of
$4\pi G \rho_c$) for the polytrope $n=3$ with $\gamma = 5/3$, $\ell=2$
and various radial orders\footnote{To obtain eigenfrequencies in the
same units than in the homogeneous case, we should multiply $\lambda^2$
by $x_1/q \simeq 162.547$; i.e. we have $\omega^2 = (x_1/q) \lambda^2$.}. As
in the homogeneous case, the anelastic approximation appears to be superior
to the subseismic one. This is illustrated by Fig.~\ref{fig2} where
we show that the anelastic relative errors are always smaller than the
subseismic ones. Still for $k=30$, the agreement with the true eigenvalue
is of about five times better with the anelastic approximation than with
the subseismic one.

We note, however, that this accuracy difference tends to be less
pronounced than for the homogeneous model. In fact, the comparison of
Fig.~\ref{fig1} and \ref{fig2} shows that the subseismic relative
errors are the same in both cases whereas the anelastic ones increase
in the polytropic case. Therefore dealing with a model with important
density variations attenuates the differences between the two
approximations.

\section{Conclusion}

We studied the properties of the subseismic and anelastic approximations
devised for low-frequency g-modes of stars.  These approximations both
assume that the Eulerian perturbations of pressure do not contribute to
their Lagrangian parts. As a consequence, acoustic waves are filtered
out. However, the equation of mass conservation differs
since the subseismic approximation keeps the contribution coming from the
Eulerian density fluctuation. We showed that this is incorrect because
it is inconsistent with the neglect of the Eulerian pressure perturbation.

As an illustration we first applied both approximations to the
low-frequency g$^-$-modes of the homogeneous star. In this case, we
found an analytic expression for the subseismic and anelastic
eigenfrequencies using power-series solutions.  We thus showed that the
anelastic approximation is more precise than the subseismic one as
expected.

As a second test, we considered the low-frequency g$^+$-modes of the
polytrope $n=3$. No analytic solutions can be found for this model
and eigenfrequencies have been computed numerically using an iterative
solver.  We thus computed the subseismic and anelastic eigenfrequencies
and compared them to the exact ones. As in the homogeneous case, the
anelastic approximation is more accurate than the subseismic one.

We also showed that Cowling's approximation is a necessary consequence
of the anelastic approximation; it is therefore unnecessary to enforce
this approximation since perturbations of self-gravity will automically
decouple from the set of equations and will not influence the
eigenfrequencies.

As already mentioned in the introduction, an obvious application of the
anelastic approximation concerns the low-frequency oscillations of
rapidly rotating stars. Applying this approximation indeed removes the
acoustic quantities and dramatically decreases the size of the
numerical problem. An example is given in Paper I with the study of the
oscillations of a rapidly rotating $\gamma$ Doradus-type star, for
which periods of oscillations are of the same order as that of rotation
(i.e. around one day).

\section*{acknowledgements}
BD has been supported by an ATER position at Universit\'e Paul Sabatier
and now by the European Commission under Marie-Curie grant no.
HPMF-CT-1999-00411 which are gratefully acknowledged.

\appendix

\section{Analytic expressions for the homogeneous star model}

\subsection{The subseismic case}

We start from Eqs.~\eq{eq7} and \eq{eq_sub}

\greq
\disp \sigma^2 \na\times \vxi = \na\times (N^2 \xi_r \er), \\ \\
\disp \na\cdot \vxi = \frac{g}{c^2} \xi_r,
\egreq

\noi where $g,c^2$ and $N^2$ are given by

\begin{eqnarray*}
& & g(r) = \frac{4\pi}{3} G \rho r, \qquad c^2 = \gamma \frac{2\pi}{3} G
\rho R^2 \lp 1 - \frac{r^2}{R^2} \rp, \\ \\
& & N^2(r) = \frac{g}{\gamma} \dnr{\Ln P} = - \frac{g^2}{c^2}.
\end{eqnarray*}

\noi The star radius is the length scale and the dynamical time
$T_{\hbox{\scriptsize dyn}} = (R^3/GM)^{1/2}$ is the time scale. In addition,
we expand the spheroidal eigenvectors $\vxi$ on spherical harmonics
as (Rieutord 1987)

\[
\vxi = \sum_{\ell=0}^{+\infty}\sum_{m=-\ell}^{+\ell} \frac{\uu(r)}{r^2}
\YL(\theta,\phi) \er + \frac{\vv(r)}{r} \na\YL,
\]

\noi where $\YL(\theta,\phi)$ denotes the normalized spherical
harmonics. We thus obtain the same radial equations coupling $\uu(r)$
and $\vv(r)$ as De Boeck et al. (1992), namely

\greq
\disp \uu - r^2 \dnr{\vv} = \frac{N^2}{\omega^2} \uu, \\ \\
\disp \dnr{\uu} - \frac{g}{c^2} \uu - \Lambda \vv = 0,
\egreq

\noi where $\Lambda = \llp$ and

\[
N^2 = - \frac{2}{\gamma} \frac{r^2}{1-r^2}, \qquad \frac{g}{c^2} =
\frac{2}{\gamma} \frac{r}{1-r^2}.
\]

\noi Eliminating $\vv$ leads to the equation for $\uu$ alone (for
clarity, we drop the subscripts $\ell$ and $m$)

\[
\begin{array}{l}
\disp \Lambda u - r^2 \ddnr{u} + \frac{2}{\gamma} \frac{r^3}{1-r^2} 
\dnr{u} \\ \\
\disp \qquad + \frac{2}{\gamma} \frac{r^2(1+r^2)}{(1-r^2)^2} u + 
\frac{2}{\gamma} \frac{\Lambda}{\omega^2} \frac{r^2}{1-r^2} u = 0.
\end{array}
\]

\noi After some algebra, this last equation may be written

\[
\begin{array}{l}
\disp \Lambda u + A r^2 u + B r^4 u + \frac{2}{\gamma} r^3 \dnr{u} \\ \\
\disp \qquad - \frac{2}{\gamma} r^5 \dnr{u} - r^2 \ddnr{u} + 
2 r^4 \ddnr{u} - r^6 \ddnr{u} = 0,
\end{array}
\]

\noi where

\[
A = \frac{2}{\gamma} \lp 1 + \frac{\Lambda}{\omega^2} \rp
-2\Lambda, \qquad B = \frac{2}{\gamma} \lp 1 - 
\frac{\Lambda}{\omega^2} \rp + \Lambda.
\]

We now expand $u(r)$ in a power-series of $r$ such as 

\[
u(r) = \sum^\infty_q c_q r^q, \qquad (q = \ell+1,\ell+3,\dots),
\]

\noi where we took into account that $u(r) = r^2 \xi_r \propto r^{\ell+1}$
near the centre. We thus obtain the following difference equation for the
$c_q$ coefficients

\[
\begin{array}{l}
\disp \lc B - \frac{2}{\gamma} q - q(q-1) \rc c_q \\ \\
\disp \qquad + \lc A + \frac{2}{\gamma} (q+2) + 2(q+2)(q+1) 
\rc c_{q+2} \\ \\
\disp \qquad + \lc \Lambda - (q+4)(q+3) \rc c_{q+4} = 0.
\end{array}
\]

\noi The convergence of the series implies that

\[
B - \frac{2}{\gamma} q - q(q-1) = 0,
\]

\noi which leads to the following expression of the subseismic
eigenfrequencies

\[
\omega^2_{\hbox{\scriptsize sub}} = \frac{2}{\gamma} \frac{\Lambda}{\Lambda -
(q-1)\lp q+\frac{2}{\gamma} \rp}.
\]

\noi In order to have an easy connection with the eigenfrequencies of
the unapproximated case, we adopt $q = \ell + 1 + 2k'$ and finally obtain

\[
\omega^2_{\hbox{\scriptsize sub}} = \frac{2}{\gamma} \frac{\Lambda}{\Lambda
- (\ell+2k')\lp \ell + 2k' +1 + \frac{2}{\gamma} \rp},
\]

\noi with $(k'=1,2,\dots)$.

\subsection{The anelastic case}

We start now from Eqs.~\eq{eq7} and \eq{eq_anel}

\greq
\disp \sigma^2 \na\times \vxi = \na\times (N^2 \xi_r \er), \\ \\
\disp \na\cdot \vxi = - \dnr{\Ln \rho} \xi_r = 0,
\egreq

\noi where we used the fact that $\rho$ is a constant for the homogeneous
star. Thus, by applying the same formalism as above, we obtain the following
anelastic radial equations for $u$ and $v$

\greq
\disp u - r^2 \dnr{v} = \frac{N^2}{\omega^2} u, \\ \\
\disp \dnr{u} - \Lambda v = 0,
\egreq

\noi and the equation for $u$ alone now reads

\[
\Lambda u - \Lambda r^2 u \lp 1 - \frac{2}{\gamma \omega^2} \rp - r^2
\ddnr{u} + r^4 \ddnr{u} = 0.
\]

As in the subseismic case, we look for a power-serie
solution $\disp u(r)= \sum^\infty_{q} c_q r^q$ which leads to the following
difference equation

\[
\begin{array}{l}
\disp \lc q(q-1) - \Lambda \lp 1 - \frac{2}{\gamma\omega^2} \rp 
\rc c_q \\ \\ 
\disp \qquad + [\Lambda - (q+2)(q+1)] c_{q+2} = 0.
\end{array}
\]

\noi The convergence of the serie requires that

\[
q(q-1) - \Lambda \lp 1 - \frac{2}{\gamma \omega^2} \rp = 0,
\]

\noi and we deduce the anelastic eigenfrequencies of the homogeneous
star as

\[
\omega^2_{\hbox{\scriptsize anel}} = \frac{2}{\gamma} \frac{\Lambda}{\Lambda 
- q(q-1)}.
\]

\noi As before, we adopt $q=\ell+1+2k'$ and obtain

\[
\omega^2_{\hbox{\scriptsize anel}} = \frac{2}{\gamma} \frac{\Lambda}{\Lambda 
- (\ell+2k')(\ell+2k'+1)},
\]

\noi with $(k'=1,2,\dots)$.

\section{The eigenvalue problem for the polytrope $n=3$}

In this appendix, we formulate the oscillation equations as a
generalized eigenvalue problem and discuss the `degeneracy' of these
equations at the star surface when the subseismic or anelastic
approximation is applied.

\subsection{The complete case}

The governing equations we need to solve are Eqs.~(1-4) with boundary
conditions (6-8). The aim is to derive a generalized eigenvalue problem
of the form

\[
{\cal M}_A \vec{\Xi} = \sigma^2 {\cal M}_B \vec{\Xi},
\]

\noi where ${\cal M}_A$ and ${\cal M}_B$ denote two differential
operators.  Here $\vec{\Xi}$ is the eigenvector associated with the
eigenvalue $\sigma^2$; it may read, for instance,

\[
\vec{\Xi} = \left| \begin{array}{l}
\rho' \\ 
\delta P \\
\xi_r \\
\xi_h \\
\phi'
\end{array} \right.
\]

\noi These equations are clearly not well adapted for an eigenvalue
problem formulation since Eqs.~\eq{eq1},\eq{eq3} and \eq{eq4} lead to
three lines of zeros in the matrix ${\cal M}_B$ making the system not
well-conditioned for iterative solvers.

We therefore prefer to use the oscillation equations of Pekeris (1938)
who obtained, after judicious substitutions, the following reduced
system (its equations 12,14 and 15; see also Hurley, Roberts \& Wright
1966)

\begin{eqnarray}
& & \disp \sigma^2 w = g \dnr{w} + \dnr{g} w - c^2 \dnr{X} - 
g(1-\gamma) X - \dnr{\Psi}, \\ \nonumber \\
& & \disp \sigma^2 \lp X - \dnr{w} - 2 \frac{w}{r} \rp = \frac{\Lambda}
{r^2} (c^2 X - g w + \Psi), \\ \nonumber \\
& & \disp \ddnr{\Psi} + \frac{2}{r}\dnr{\Psi} - \Lambda \frac{\Psi}{r^2}
- 4 \pi G \lp w \dnr{\rho} + \rho X \rp = 0,
\end{eqnarray}

\noi where $w = \vi \cdot \er$ (should not be confused with the
dimensionless eigenfrequencies $\omega$), $X=\na\cdot \vi$ and $\Psi =
i \sigma \psi$, $\psi$ being the gravitational potential perturbation.

In order to obtain the simplest dimensionless polytropic equations, we
choose two new length and time scales. As length scale, we take $R/x_1$
whereas $T_0 = (4 \pi G \rho_c)^{-1/2}$ is our time scale. Here $x_1$
is related to $P_c$ and $\rho_c$ by (see e.g. Hansen \& Kawaler 1994)

\[
\lp \frac{R}{x_1} \rp^2 = \frac{n+1}{4\pi G} \frac{P_c}{\rho^2_c},
\]

\noi where $P_c$ and $\rho_c$ respectively denote the central pressure
and density of the polytrope. With these scales, we have for example
the following relations

\[ \begin{array}{l}
\disp \sigma^2 = 4 \pi G \rho_c \lambda^2 = \frac{x_1}{q} \frac{GM}{R^3}
\omega^2, \qquad g=- \frac{R}{x_1} 4 \pi G \rho_c \dnx{\theta}, \\ \\
\disp c^2 = \lp \frac{R}{x_1} \rp^2 4 \pi G \rho_c \frac{\gamma}{n+1} 
\theta, \qquad q=- \lp \dnx{\theta} \rp_{x=x_1}.
\end{array}
\]

\noi We note that for the polytrope $n=3$, $x_1 \simeq 6.89685$ and $q
\simeq 4.24297 \times 10^{-2}$. Moreover, we recall that the function
$\theta (x)$ is solution of the Lane-Emden equation

\beq
\frac{1}{x^2} \dnx{} \lp x^2 \dnx{\theta} \rp = - \theta^n, \qquad
\theta(0) = 1, \qquad \theta (x_1) = 0.
\eeqn{lane}

We thus obtain the dimensionless system (the prime is for the derivative
$d/dx$)

\greq
\disp \lambda^2 w = - \theta' w' + \lp \theta^n + 2
\frac{\theta'}{x} \rp w - \frac{\gamma}{n+1} \theta X' \nonumber \\ 
\nonumber \\
\hspace{1cm} \disp +(1-\gamma) \theta' X - \Psi', \nonumber \\ \nonumber \\
\disp \lambda^2 \frac{x^2}{\Lambda} \lp X - w' - \frac{2}{x} w \rp =
\theta' w + \frac{\gamma}{n+1} \theta X + \Psi, \nonumber \\ \nonumber \\
\disp \Psi'' + \frac{2}{x} \Psi' - \frac{\Lambda}{x^2} \Psi - n \theta'
\theta^{n-1} w - \theta^n X = 0, \label{s1.a2}
\egreq

\noi with the following set of boundary conditions

\greq
w = \Psi = 0 \ltex{at} x=0, \\ \\
\disp \Psi' + \frac{\ell+1}{x_1} \Psi = 0 \ltex{at} x=x_1.
\label{s2.a2}
\egreq
 
Systems \eq{s1.a2} and \eq{s2.a2} may still be formally written as

\beq
{\cal M}_A \vec{\Xi} = \lambda^2 {\cal M}_B \vec{\Xi},
\eeqn{gegv}

\noi where ${\cal M}_A$ and ${\cal M}_B$ denote two differential operators and
$\lambda^2$ are the real eigenvalues associated with the eigenvectors
$\vec{\Xi}$ such as

\[
\vec{\Xi} = \left| \begin{array}{l}
w \\
X \\
\Psi
\end{array} \right.
\]

\subsection{The subseismic and anelastic cases}

We have shown in Section 3.1 that the Cowling approximation is
necessary when the subseismic or anelastic approximation are used; thus
we now have to deal with

\begin{eqnarray}
& & \disp \sigma^2 \vxi = \na \lp \frac{P'}{\rho} \rp + N^2 \xi_r \er =
\na \chi + N^2 \xi_r \er, \label{eq1.a2} \\ \nonumber \\
& & \disp \na\cdot \vxi = \frac{g}{c^2} \xi_r \ltex{or} \na\cdot \vxi = 
- \dnr{\Ln \rho} \xi_r, \label{eq2.a2}
\end{eqnarray}

\noi where we defined $\chi = P' / \rho$. By using the same scales as above
and eliminating $v$ with Eq.~\eq{eq2.a2}, we obtain the following
dimensionless system for $u$ and $p$

\begin{eqnarray}
& & \disp \lambda^2 u = x^2 p' + \lp n - \frac{n+1}{\gamma} \rp 
\frac{\theta'^2}{\theta} u, \label{eq3.a2} \\ \nonumber \\
& & \disp \lambda^2 \lp u' + \alpha \frac{\theta'}{\theta} u \rp = \Lambda p.
\label{eq4.a2}
\end{eqnarray}

\noi Here the parameter $\alpha$ differs according to the
approximation used, namely

\[
\alpha = \frac{n+1}{\gamma}: \hbox{ subseismic}, \qquad \alpha =
\frac{n}{\gamma}: \hbox{ anelastic},
\]

\noi whereas $p \equiv p^\ell_m$ is related to the projection of $\chi$
on the spherical harmonics as

\[
\chi = \sum_{\ell=0}^{+\infty}\sum_{m=-\ell}^{+\ell} p^\ell_m
\YL(\theta,\phi).
\]

At the surface, we have $\theta (x_1) = 0$ thus regular singularities
appear in Eqs.~\eq{eq3.a2} and \eq{eq4.a2}. To remove them, a solution
would consist in multiplying these two equations by $\theta$ to obtain
the new system

\begin{eqnarray}
& & \disp \lambda^2 \theta u = \theta x^2 p' + \lp n - \frac{n+1}{\gamma} \rp 
\theta'^2 u, \\ \nonumber \\
& & \disp \lambda^2 \lp \theta u' + \alpha \theta' u \rp = \Lambda \theta p.
\end{eqnarray}

\noi The singularities indeed disappear but are now replaced by a
surface degeneracy since each equation gives the same solution $u=0$ at
$x=x_1$ (which is however physically correct when using the subseismic
or anelastic approximation; see Section 3.1). This is however making the
discretized eigenvalue problem singular and this should be avoided.

Following Hurley et al. (1966), we preferred to use the Frobenius
method. We first develop $\theta(x)$ near $x=x_1$ as

\[
\theta(x) = \theta(x_1) + (x-x_1) \theta'(x_1) + \od{x-x_1}^2,
\]

\noi and, as $\theta(x_1) = 0$, 

\beq
\frac{\theta'}{\theta} \sim \frac{1}{x-x_1} \ltex{for} x \rightarrow x_1.
\eeq

\noi We next expand $u(x)$ and $p(x)$ in series of the form

\[
u(x) = \sum^{+\infty}_{q=0} a_q (x-x_1)^q, \qquad p(x) = 
\sum^{+\infty}_{q=0} b_q (x-x_1)^q,
\]

\noi and obtain, using Einstein's notations,

\[
u' = q a_q (x-x_1)^{q-1}.
\]

\noi Thus Eq.~\eq{eq4.a2} reads now

\[
\lambda^2 \lc q a_q (x-x_1)^{q-1} + \alpha a_q (x-x_1)^{q-1} \rc =
\Lambda b_q (x-x_1)^q.
\]

\noi Equating the lowest power of $(x-x_1)$, which is here
$(x-x_1)^{-1}$, gives again

\[
a_0 = 0\quad \Leftrightarrow \quad u = 0,
\]

\noi whereas the next power of $(x-x_1)$ gives

\[
\lambda^2 ( a_1 + \alpha a_1 ) = \Lambda b_1 \quad \Leftrightarrow 
\quad \lambda^2 ( 1 + \alpha ) u' = \Lambda p.
\]

We therefore obtain the second boundary condition and the system we
need to solve is

\greq
\disp \lambda^2 u = x^2 p' + \lp n - \frac{n+1}{\gamma} \rp
\frac{\theta'^2}{\theta} u, \\ \\
\disp \lambda^2 \lp u' + \alpha \frac{\theta'}{\theta} u \rp = \Lambda
p, \\ \\
u = 0 \ltex{and} \lambda^2 ( 1 + \alpha ) u' = \Lambda p \ltex{at} x=x_1.
\egreq

\noi which can be rewritten as \eq{gegv} with

\[
\vec{\Xi} = \left| \begin{array}{l}
u \\
p \\
\end{array} \right.
\]

\label{lastpage}

\end{document}